\documentstyle[aps,twocolumn,epsfig]{revtex}
\draft
\newcommand{\<}{\langle}
\renewcommand{\>}{\rangle}

\newcommand{\be}{\begin{equation}}
\newcommand{\ee}{\end{equation}}
\newcommand{\bea}{\begin{eqnarray}}
\newcommand{\eea}{\end{eqnarray}}

\newcommand{\mtwo}[4] {\left( \matrix{{#1}&{#2}\cr{#3}&{#4}} \right)}

\begin{document}
\wideabs{


\title{Realization of quantum process tomography in NMR} 

\author{Andrew M. Childs,$^{1,2,3}$ Isaac L. Chuang,$^1$ 
        and Debbie W. Leung$^{1,4,5}$}

\address{
	{$^1$ IBM Almaden Research Center, San Jose, CA 95120} \\
	{$^2$ Physics Department, California Institute of Technology,
	Pasadena, CA 91125} \\
	{$^3$ Center for Theoretical Physics, Massachusetts Institute of
	Technology, Cambridge, MA 02139} \\
	{$^4$ Quantum Entanglement Project, ICORP, JST, Edward Ginzton
	Laboratory, Stanford University, Stanford, CA 94305} \\
	{$^5$ IBM T. J. Watson Research Center, Yorktown Heights, NY 10598} }

\date{6 December 2000}

\maketitle


\begin{abstract}
Quantum process tomography is a procedure by which the unknown dynamical
evolution of an open quantum system can be fully experimentally
characterized.  We demonstrate explicitly how this procedure can be
implemented with a nuclear magnetic resonance quantum computer.  This allows
us to measure the fidelity of a controlled-{\sc not} logic gate and to
experimentally investigate the error model for our computer.  Based on the
latter analysis, we test an important assumption underlying nearly all
models of quantum error correction, the independence of errors on different
qubits.
\end{abstract}



} 
\narrowtext

\section{Introduction}

Experimental characterization of the dynamical behavior of open quantum
systems has traditionally revolved around semiclassical concepts such as
coupling strengths, relaxation rates, and phase coherence
times~\cite{Louisell73a,Gardiner91}.  However, quantum information theory
tells us that so few parameters barely even begin to represent the full
dynamics which quantum systems are capable of; for example, the evolution
between two fixed times of a two-level quantum system (a qubit), coupled to
an arbitrary reservoir, is described by {\em twelve} real
parameters~\cite{Chuang97a}.  For a two qubit system, this number grows to
$240$, and in general, for $n$ qubits it is $16^n - 4^n$.

Of course, not all of these parameters are relevant for every physical
process, and understanding the physics can boil down this huge parameter
space to just a few important numbers.  On the other hand, when the physical
origins of a quantum process are {\em not} understood, or in question, it is
invaluable to know that all these parameters can, in principle, be
experimentally measured --- albeit with exponential effort (in $n$) ---
using a procedure known as {\em quantum process tomography}.  This method is
a direct quantum extension of the classical concept of ``system
identification'' which enables control of dynamical systems, and provides
powerful measurement techniques for characterizing unknown systems.  For
closed systems, process tomography is the analog of determining the truth
table of an unknown function; quantum-mechanically, this means the unitary
transform involved in a quantum computation.

The procedure for quantum process tomography has been described in detail in
the literature~\cite{Chuang97a,Poyatos97a,Leung}.  It relies upon the
ability to prepare a complete set of quantum states ${\rho_j}$ as input to
the unknown process ${\cal E}_t$, and the ability to measure the density
matrices of the output quantum states from the process ${\cal E}_t(\rho)$.
As a result, one obtains a complete ``black box'' characterization of ${\cal
E}_t$.  By systematically repeating this procedure for increasing process
time $t$, one can also obtain a complete master equation description of the
process~\cite{Buzek98a}.

This procedure seems to provide an ideal way to characterize and test
quantum gates on a small number of qubits --- for example, those realized
using nuclear magnetic resonance (NMR)
techniques~\cite{Gershenfeld97a,Cory97a}.  However, to date, this has not
been practiced because of two problems: first, it is desirable to explicitly
include the maximally mixed state as one of the $\rho_j$'s, so as to
directly subtract its contribution, but such a state cannot be prepared by
unitary actions alone.  Second, one can only measure traceless observables,
and thus one cannot obtain the entire output density matrix of a process.

Here, we show explicitly how these hurdles can be overcome, and demonstrate
the use of these methods in addressing two outstanding issues in quantum
computation: the fidelity of a quantum logic gate, the controlled-{\sc not}
gate, and the validity of the independent error model which is widely assumed
for quantum error correction and fault tolerant computation.

The paper is divided into two major sections, which describe the theory, and
our experiment.  We begin with a review of quantum process tomography (QPT),
then we explain our extensions to the basic procedure which enable QPT with
NMR, and present theoretical models for decoherence in NMR.  We then
describe our experiment and procedure, and our two main experimental
results.

\section{Theory}

\subsection{Quantum process tomography}

Quantum process tomography, as introduced by Chuang and
Nielsen~\cite{Chuang97a}, can be summarized as follows.  A general quantum
operation on a quantum state $\rho$ is a superoperator $\cal E$, a linear,
trace-preserving, completely positive
map~\cite{Schumacher96,Choi75a,Kraus83a}.  We consider only the case where
$\cal E(\rho)$ has the same dimension as $\rho$.
A common form for $\cal E$ is the operator-sum representation,
\be
  {\cal E}(\rho) = \sum_k A_k \rho A_k^\dagger
\,,
\label{eq:osr}
\ee
where $\sum_k A_k^\dagger A_k = I$.  One can easily demonstrate the
transformation to a fixed-basis expansion of the quantum operation such that
the information about the process lies in a set of coefficients $\chi_{mn}$
instead of a set of operators.  In this form, $\cal E$ maps an initial
density matrix $\rho$ according to
\be
  {\cal E}(\rho) = \sum_{m,n} \chi_{mn} A_m \rho A_n^\dagger
\,.
\ee
In this expression, the $A_m$ are a basis for operators on the space of
density matrices, and $\chi$ is a positive Hermitian matrix with elements
$\chi_{mn}$.  In the following, we will omit explicit summations and adopt
the convention that repeated indices are summed over.  Because of the
restriction that $\cal E$ must preserve the trace of $\rho$, $\chi$ contains
$N^4-N^2$ independent parameters for an $N$ dimensional system (for $n$
spins, $N=2^n$).

Let the $N^2$ matrices $\rho_j$ be a basis for density matrices.  Applying
$\cal E$ to this basis gives
\be
  {\cal E}(\rho_j) = \lambda_{jk} \rho_k
\,.
\label{eq:linsuper}
\ee
Using quantum state tomography~\cite{VR89}, one can experimentally determine
$\lambda_{jk}$, which fully specifies $\cal E$.  To determine $\chi$ from
$\lambda$, we define $\beta$ by
\be
  A_m \rho_j A_n^\dagger = \beta^{mn}_{jk} \rho_k
\,,
\ee
which is fully specified by the choice of $A_m$ and $\rho_j$.  Then one can
show that
\be
  \beta^{mn}_{jk}\chi_{mn} = \lambda_{jk}
\,.
\ee
We may think of $\beta$ as a matrix and $\lambda$ and $\chi$ as vectors, with
$mn$ a composite column index and $jk$ a composite row index; then $\beta
\vec\chi = \vec\lambda$.  Using the pseudoinverse $\kappa$ of $\beta$
(also called the Moore-Penrose generalized inverse), which may be computed
independent of measurement results, we have
\be
  \vec\chi = \kappa \vec\lambda
\,.
\ee

\subsection{QPT in NMR}
\label{sec:nmrqpt}

Implementing QPT in high temperature NMR presents two main obstacles.
First, the part of the system which can be manipulated by unitary operations
is but a small deviation from a maximally mixed state.  Second, one can only
measure traceless observables, and the scaling of these observables with
respect to the traceful part of the system is not immediately known.
Fortunately, both of these hurdles can be overcome.

The Hamiltonian for solution NMR is well approximated by
\be
  H=-\hbar \omega_j {Z_j \over 2} + h J_{jk} {Z_j Z_k \over 4}
\,,
\label{nmrhamiltonian}
\ee
where
\be
  Z_j = \left( \matrix{1 & 0 \cr 0 & -1} \right)
\ee
is the Pauli operator for the $j$th spin along the direction of the magnetic
field, $\omega_j$ is its Larmor frequency, and the sign of the first term
is chosen so that the ground state is spin up.  The second term represents a
small scalar coupling between spins known as the $J$-coupling.

The initial state of an NMR quantum computer is the thermal equilibrium
state with density matrix
\be
  \rho_\infty = {e^{-H/k_BT} \over {\rm tr}(e^{-H/k_BT})} = c I + \Delta_\infty
\,,
\ee
where $c=2^{-n}$ for a system of $n$ spins (so that ${\rm
tr}(\rho_\infty)=1$) and $\Delta_\infty$ is traceless.  In general, we refer
to the traceless part of any density matrix $\rho$ as the deviation density
matrix.  If we neglect the small coupling between spins and consider high
temperature with respect to the Larmor frequencies, the equilibrium density
matrix is approximately
\be
  \rho_\infty \approx c\left(I+{\hbar\omega_j Z_j \over 2 k_B T}\right)
\,.
\ee

By performing unitary operations on the system, we may rotate the $\omega_j
Z_j$ deviation, allowing us to prepare a linearly independent set of inputs
$\rho_j$.  Because all of the interesting information will lie in this
deviation, we would like to be able to prepare an input with no deviation
(i.e., the maximally mixed state), so that we may directly subtract the
contribution of the identity.  Such an input clearly cannot be prepared by
unitary means.  However, the maximally mixed state {\em can} be prepared
from the thermal state by applying a $\pi \over 2$ pulse followed by a 
uniformly spatially varying RF pulse.  Different parts of the sample
experience different amounts of rotation, so that the ensemble is
effectively dephased to the maximally mixed state.

The second difficulty can be handled by calibrating measurements to the
known initial state of the system.  By state tomography, we can measure
$\tilde\Delta = d\Delta_\infty$, where the constant $d$ is arbitrary --- it
depends in detail on the gain of the amplifiers, the efficiency of the RF
coils, etc.  But the initial deviation density matrix is known to be
$c{\hbar\omega_j Z_j \over 2 k_B T}$.  Hence we may calibrate our apparatus
by determining the value $d$ that makes the measured $\tilde\Delta/d$ closest
to the theoretical $\Delta_\infty$.  In practice, we effect this calibration
by normalizing all measurements (integrals of spectral peaks) for spin $j$
to a reference measurement taken on that spin (a $\pi \over 2$ pulse applied
to the thermal state) multiplied by $c{\hbar \omega_j \over 2 k_B T}$.

\subsection{Decoherence primitives}

Decoherence of a single NMR spin in solution is generally characterized by
two well-known processes, amplitude damping and phase damping.  Here, we
present these processes, as well as generalizations to multiple spins.  To
facilitate the description of multiple simultaneous processes, we describe
amplitude damping and phase damping as semigroups in terms of their
generators.

Assuming that a quantum operation can be described by a norm continuous
one-parameter semigroup, given a superoperator ${\cal E}_t$ satisfying the
stationarity and Markovity condition ${\cal E}_s {\cal E}_t={\cal
E}_{s+t}$~\cite{MultNotation}, its generator $\cal Z$ is defined
as~\cite{Davies} 
\be
\label{eq:generator}
  {\cal Z}(\rho)=\lim_{t \to 0} {{\cal E}_t(\rho)-\rho \over t}
\,.
\ee
The generators have the advantage that they easily describe the simultaneous
action of multiple processes.  In particular, we have the Trotter
formula~\cite{Davies,Trotter},
\be
  {\cal E}^{(1+2)}_t = \lim_{n \to \infty} 
    \left({\cal E}^{(1)}_{t/n} {\cal E}^{(2)}_{t/n} \right)^n
\,,
\ee
where
\be
  {\cal Z}^{(1+2)} = {\cal Z}^{(1)} + {\cal Z}^{(2)}
\,,
\ee
with obvious generalization to more than two processes.

Furthermore, we may return to the superoperator form by exponentiating the
generator~\cite{Davies}:
\bea
\label{eq:exp}
  {\cal E} &=& \lim_{n \to \infty} 
               \left(1-{t \over n}{\cal Z}\right)^{-n} \\
           &=& e^{{\cal Z}t}
\,.
\eea
This exponentiation may be defined by its Taylor series in the case of
interest where $\cal Z$ is bounded.  To implement exponentiation in
practice, it is useful to work with a manifestly linear representation,
i.e., that of Eq.~(\ref{eq:linsuper}).  In this case, a superoperator and
its generator can be represented as matrices which have the property that
composition of superoperators corresponds to matrix multiplication.  Then
the procedure of Eq.~(\ref{eq:exp}) corresponds to matrix exponentiation.

We now consider the single-qubit versions of phase and amplitude damping.
For further discussion of these processes, including operator-sum
representations, see~\cite{MikeAndIke}.  We write the density matrix of a
qubit as
\be
  \rho = \mtwo{\rho_{00}}{\rho_{01}}{\rho_{10}}{\rho_{11}}
\,.
\ee

Phase damping can be thought of as a consequence of random phase kicks.  It
acts on a qubit as \be
  {\cal E}^{\rm PD}_t(\rho)
     = \mtwo{\rho_{00}}{e^{-\gamma t}\rho_{01}}
            {e^{-\gamma t}\rho_{10}}{\rho_{11}}
\,,
\ee
for some damping rate $\gamma$; i.e., it has a generator which acts as
\be
  {\cal Z}^{\rm PD}(\rho) 
    = -\gamma \left(\matrix{0 & \rho_{01} \cr \rho_{10} & 0}\right)
\,.
\ee
Ordering the matrix elements of $\rho$ in the vector $(\rho_{00}, \rho_{01},
\rho_{10}, \rho_{11})^T$ (which effectively defines the basis used in
Eq.~(\ref{eq:linsuper})), we have
\be
{\cal Z}^{\rm PD}=\left(\matrix{ 0 & 0       & 0       & 0 \cr
                                 0 & -\gamma & 0       & 0 \cr
                                 0 & 0       & -\gamma & 0 \cr
                                 0 & 0       & 0       & 0 }\right)
\,.
\ee

Generalized amplitude damping is a process whereby the qubit may exchange
energy with a reservoir at some fixed temperature.  It acts on a qubit as
\be
  {\cal E}^{\rm GAD}_t(\rho) =
    \mtwo{k_1 \rho_{00} + k_2 \rho_{11}}
         {e^{-\Gamma t/2} \rho_{01}}
         {e^{-\Gamma t/2} \rho_{10}}
         {k_3 \rho_{00} + k_4 \rho_{11}}
\,,
\ee
where $k_2 \equiv (1-\bar n)(1-e^{-\Gamma t})$, $k_3 \equiv \bar
n(1-e^{-\Gamma t})$, $k_1 \equiv 1-k_3$, and $k_4 \equiv 1-k_2$, with
$\Gamma$ a damping rate and $\bar n$ a temperature parameter.  Its generator
acts as
\be
{\cal Z}^{\rm GAD} = -\Gamma
  \left(\matrix{\bar n   & 0           & 0           &\bar n-1 \cr
                 0       & {1 \over 2} & 0           & 0   \cr
                 0       & 0           & {1 \over 2} & 0   \cr
                 -\bar n & 0           & 0           & 1-\bar n }\right)
\,,
\ee
Computing the fixed point of this process, we find that we may interpret it
as occuring at a temperature 
\be
  k_B T = {\Delta E \over \log{1-\bar n \over \bar n}}
\,,
\ee
where $\Delta E$ is the splitting between the ground and excited states of
the system.

Heuristically, we expect phase damping to describe the $T_2$ (dephasing)
process of NMR.  Similarly, generalized amplitude damping should describe
the $T_1$ process of NMR, whereby a spin relaxes to align with the magnetic
field.  Note from the geometry of the Bloch sphere that amplitude damping
necessarily includes loss of phase information, so it also contributes to
the $T_2$ process.  However, in typical systems, $T_1$ is much longer than
$T_2$, so the contribution of amplitude damping to dephasing is small.

The simplest extension of these processes to a multiple-spin system is to
allow the individual processes to act independently on each qubit.  However,
this is far from the most general scenario: the decoherence might be
correlated.  Although we have been unable to find a suitable correlated
generalization of amplitude damping, we have incorporated a simple model of
correlated phase damping due to Zhou into our analysis~\cite{ZhouPrivate}.
In this model, each spin experiences a random phase shift with zero mean.
The amount of damping and the extent of correlation is defined by the
covariance matrix of the phase shifts.  By explicit calculation, one may
easily find that this leads to a generator of the form
\bea
{\cal Z}^{\rm CPD}={\rm
diag}&[&0,-\gamma_2,-\gamma_1,-(\gamma_1+\gamma_2+\gamma_3), \nonumber\\
     && -\gamma_2,0,-(\gamma_1+\gamma_2-\gamma_3),-\gamma_1, \nonumber\\
     && -\gamma_1,-(\gamma_1+\gamma_2-\gamma_3),0,-\gamma_2, \nonumber\\
     && -(\gamma_1+\gamma_2+\gamma_3),-\gamma_1,-\gamma_2,0]
\,,
\eea
where $\gamma_1$ and $\gamma_2$ may be interpreted as rates for independent
phase damping on spins $1$ and $2$, and $\gamma_3$ may be interpreted as a
rate for correlated damping.  Indeed, in the case $\gamma_3=0$, this model
reproduces independent phase damping.

Our completed model has the generator
\be
\label{eq:dampmodel}
  {\cal Z}=  {\cal Z}^J + {\cal Z}^{\rm CPD}
           + {\cal Z}^{\rm GAD}_1 + {\cal Z}^{\rm GAD}_2
\,,
\ee
where a subscript indicates which spin is acted on and ${\cal Z}^J$ is the
generator corresponding to the Hamiltonian
\be
  H^J = h J {Z_1 Z_2 \over 4}
\,.
\ee
In this Hamiltonian, the Zeeman terms are dropped from
Eq.~(\ref{nmrhamiltonian}) because we work in a frame rotating at the
Larmor frequencies of the two spins.  To produce the model superoperator at
any given time, we simply exponentiate Eq.~(\ref{eq:dampmodel}).

\section{Experiment}

We have implemented QPT on a system of two spins in an NMR apparatus.  The
experiments were performed at the IBM Almaden Research Center using an
Oxford Instruments wide-bore magnet and a 500 MHz Varian Unity Inova
spectrometer with a Nalorac triple resonance HFX probe.  The sample was
approximately $0.5~{\rm mL}$ of $200$ millimolar isotopically labeled
chloroform ($^{13}{\rm CHCl}_3$) in d6-acetone, where the two qubits were
the proton (first spin) and carbon (second spin).  This volume was chosen to
make the sample fairly short, as this gave the most effective gradient pulse
characteristics.

\subsection{Results: Gate fidelities}

One application of quantum process tomography is to diagnose the accuracy of
a quantum computation.  When one applies a series of gates to perform the
unitary operation $U$, the computer will actually perform a superoperator
$\cal E$, which one hopes is close to $U$.  That closeness can be captured
by many distance measures --- for example, the minimum gate fidelity,
defined as
\be
\label{eq:fidelity}
F=\min_{|\psi\>} \<\psi| U^\dagger {\cal E}(|\psi\>\<\psi|) U |\psi\>
\,.
\ee
$F$ represents the smallest possible overlap between a state acted on by $U$
and the same state acted on by $\cal E$.  Note that minimizing over pure
states is sufficient because the fidelity is convex and because an arbitrary
density matrix can be written as a convex sum of pure states.

\begin{figure}
\begin{center}
\mbox{\psfig{file=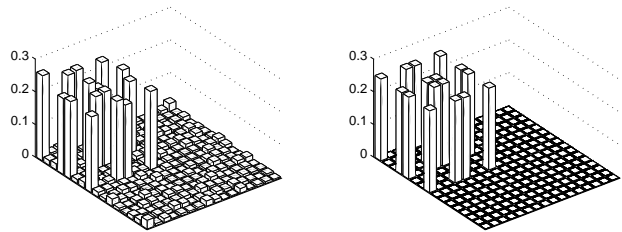,width=3.25in}}
\end{center}
\caption{$\chi$ matrices for the controlled-{\sc not} gate.  The matrix on
the left was experimentally measured by QPT, and that on the right is
theoretical.  Only the magnitudes of the matrix elements are shown; the
phases also corresponded well between theory and experiment.  The matrix
element in the far left corner connects $II \leftrightarrow II$, and the
elements are ordered as $II, XI, \ldots, YZ, ZZ$.}
\label{fig:cnot}
\end{figure}

Using QPT, we measured the minimum gate fidelity for a controlled-{\sc not}
operation, a well-known member of universal gate sets, with
\be
U_{\sc cnot} = \left(\matrix{1 & 0 & 0 & 0 \cr
                             0 & 1 & 0 & 0 \cr
			     0 & 0 & 0 & 1 \cr
			     0 & 0 & 1 & 0}\right)
\ee
The NMR pulse sequence chosen to implement this gate was
\be
\bar Y_1 X_1 Y_1 \bar Y_2 d\left({1 \over 2J}\right) Y_2 X_2
\,,
\ee
where time goes right to left, $X_j$ denotes a $\pi \over 2$ $x$ pulse on
spin $j$, a bar denotes an inverse pulse, $d(\cdot)$ denotes a time
evolution during which no pulses are applied, and $J=215~{\rm Hz}$ is the
coupling constant for the scalar $J$-coupling between the spins.  Procedures
similar to those used in~\cite{Vandersypen} were adopted to implement this
pulse sequence.  The pulse lengths were approximately $10-20~{\rm \mu s}$,
sufficiently fast that the effect of $J$-coupling could be neglected during
a pulse.  Using quantum process tomography, we determined $\chi$ for this
process, effectively determining $\cal E$.  This result is shown in
Fig.~\ref{fig:cnot}.

Given $\chi$, we may calculate gate fidelities.  For the controlled-{\sc
not} gate, numerically minimizing over input states gives $F_{\sc cnot}=0.80
\pm 0.04$.

This figure may be thought of as a rough benchmark for the quality of gates
that can be implemented using current NMR quantum computers.  However,
several caveats should be addressed.  Foremost, the preparation and readout
steps performed to do the process tomography contribute to much of the
error.  Performing process tomography on a null computation ($U=I$), we find
$F_I = 0.90 \pm 0.03$.  The primary sources of error are most likely
imperfect state preparation and imperfect state tomography due to imperfect
pulse calibration and inhomogeneity of the RF
fields~\cite{BetterResultsNote}.

Also, in a long computation, experimental results suggest that there may be a
significant cancellation of errors~\cite{Vandersypen}.  Typically, the major
contribution to the error introduced by an individual gate will be largely
due to systematic errors rather than fundamentally irreversible decoherence.
If the pulse sequence exhibits some degree of symmetry, these systematic
errors may at least partially cancel.  Thus the fidelity of a long sequence
of gates cannot be deduced from the individual gate fidelities, though one
would certainly expect it to be subadditive.

Of course, the minimum gate fidelity is a pessimistic standard; it may be
more reasonable to consider the {\em average} gate fidelity.  To numerically
calculate the average fidelity, we must be able to sample uniformly from
quantum states according to an appropriate probability measure.  Although
this problem is subtle for the case of mixed states, there is a
straightforward choice for pure states: we take the unitary transform of a
fixed state, where the unitary operator is chosen according to the Haar
measure (the unique invariant measure on a Lie group).  Using the procedure
described in ~\cite{HaarMeasure}, we calculated average fidelities with
$10^5$ randomly chosen unitaries, giving $F^{\rm avg}_{\sc cnot} = 0.955$
and $F^{\rm avg}_I = 0.960$.  Thus, the average fidelity is significantly
better than the worst case; the error rate $1-F$ improves by about one order
of magnitude.

\subsection{Results: Decoherence characterization} 
\label{sec:decoherence}

To characterize the decoherence occuring in the chloroform system, we used
QPT to measure $\chi$ over the various relevant time scales.  For an NMR
spin, the relevant scales are $1 \over 2J$, $T_1$, and
$T_2$~\cite{RotFrameNote}.  We thus sampled at times given by $({1
\over 2})^j {1\over 2\tilde J}$ for integers $j \in [0,4]$, $({1 \over 2})^j
\tilde T_1$ for $j \in [-1,4]$, and $({2 \over 3})^j \tilde T_2$ for $j \in
[0,9]$, where $\tilde J=215~{\rm Hz}$ is the known value of the
$J$-coupling, $\tilde T_1=20~{\rm s}$ is a time scale on the order of $T_1$,
and $\tilde T_2=0.5~{\rm s}$ is an intermediate time scale on the order of
$T_2$.

By executing process tomography, we are able to determine $\chi$ as a
function of time.  However, $\chi$ is a large, complex collection of numbers
that cannot be easily interpreted.  To better understand the results, we fit
the data to a model process, $\chi_m$, which hopefully provides a reasonable
description of the relevant physics.  Specifically, we use the model which
results from exponentiating Eq.~(\ref{eq:dampmodel}).

\begin{figure}
\begin{center}
\mbox{\psfig{file=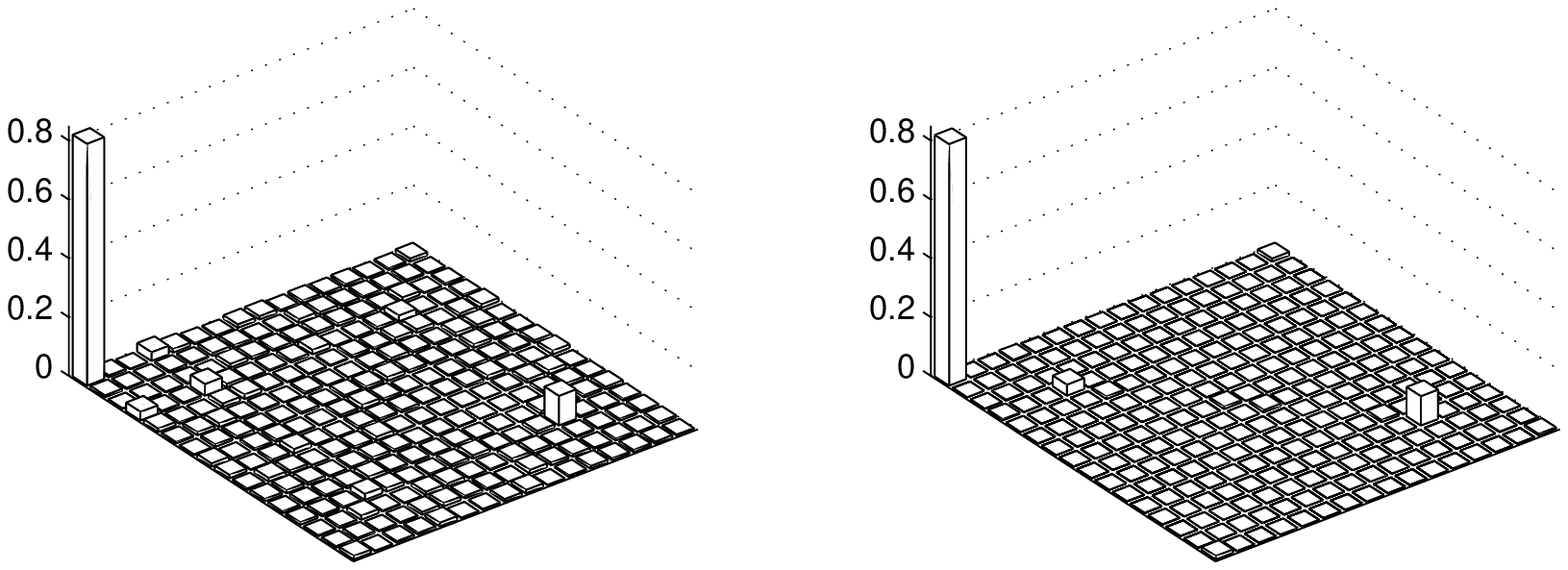,width=3.25in}}
\mbox{\psfig{file=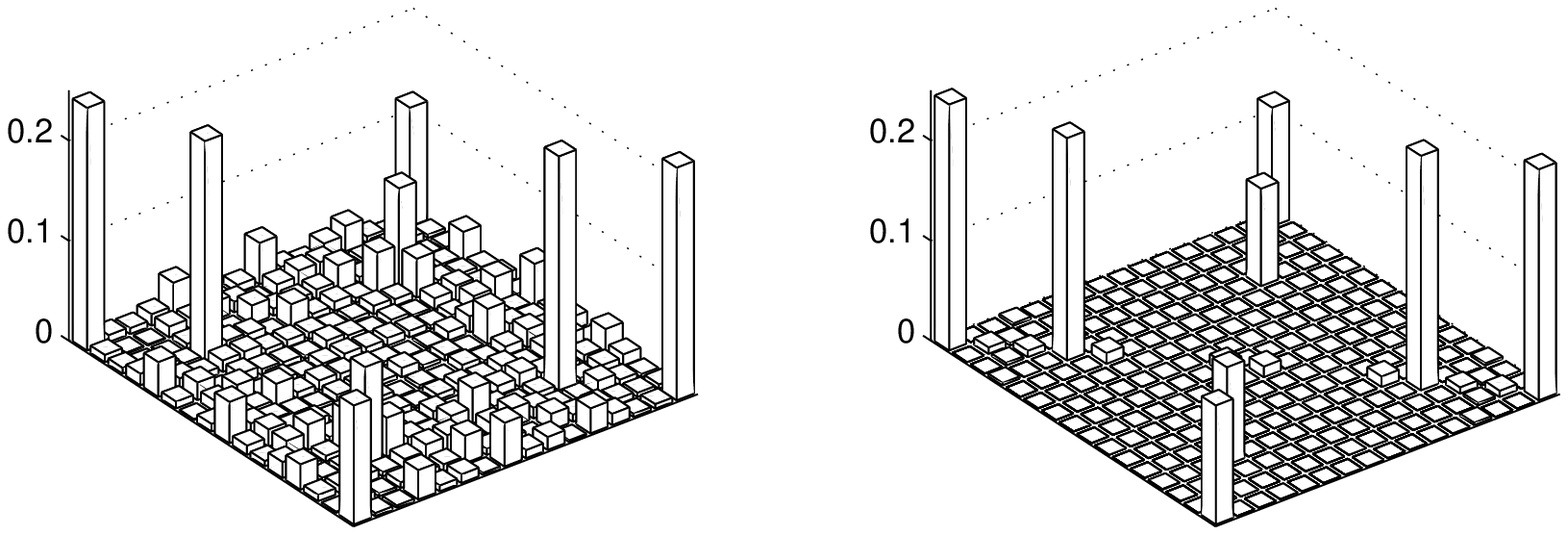,width=3.25in}}
\mbox{\psfig{file=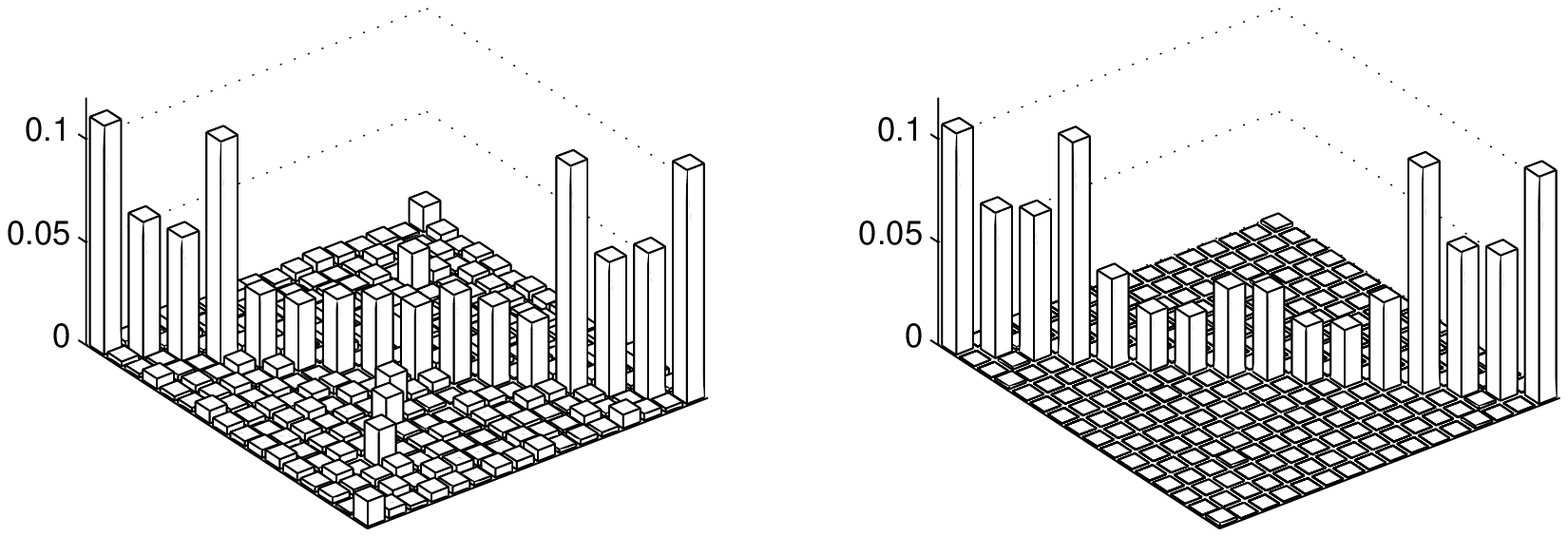,width=3.25in}}
\end{center}
\caption{$\chi$ matrices for decoherence at various times.  Matrices on the
left are experimentally measured using QPT, and those on the right are fits
to Eq.~(\ref{eq:dampmodel}).  Only the magnitudes of the matrix elements are
shown, but their phases were also described well by the model.  From top to
bottom, $t=0.065~{\rm s},\ 0.5~{\rm s},\ 20~{\rm s}$.  At $t=0.065~{\rm s}$,
the time was chosen to be an integer number of $J$-coupling periods, which
explains the lack of $IZ$, $ZI$, and $ZZ$ terms in the corners.  At
$t=0.5~{\rm s}$, note that the $IZ \leftrightarrow ZI$ terms arise from a
combination of $J$-coupling and {\em independent} phase damping, not from
correlated phase damping.  At $t=20~{\rm s}$, note that the nonzero
anti-diagonal terms found in the experiment do not appear in the fit even
though such terms may arise in our amplitude damping model.  Due to the
details of the fitting procedure, the optimal fit matches large elements
closely but does not capture the detail of the small anti-diagonal.}
\label{fig:dampchi}
\end{figure}

For each $\chi$, we determined the closest fit to our model by numerically
minimizing ${\rm tr}[(\chi_{\rm m}-\chi)^\dagger(\chi_{\rm m}-\chi)]$, where
$\chi_{\rm m}$ is a model superoperator derived from
Eq.~(\ref{eq:dampmodel}).  This figure of merit was on the order of $5\%$
for most fits.  The experimentally measured $\chi$ matrices as well as the
corresponding fits $\chi_m$ are shown in Fig.~\ref{fig:dampchi} for three
delay times.

\begin{figure}
\begin{center}
\mbox{\psfig{file=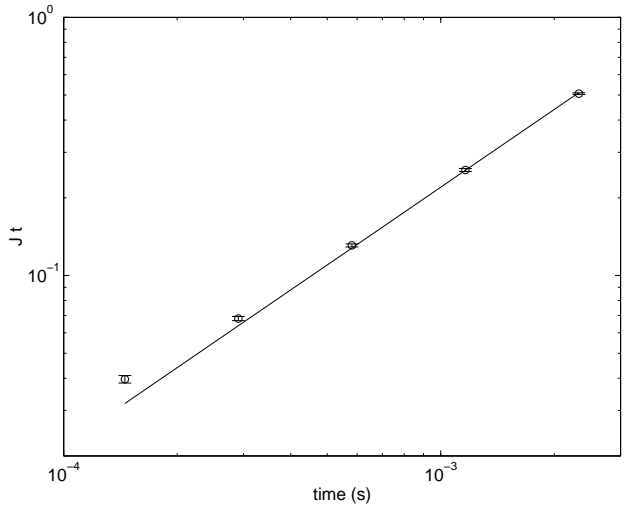,width=3.25in}}
\end{center}
\caption{Determination of the strength of the scalar $J$-coupling between
spins.}
\label{fig:jcoupling}
\end{figure}

\begin{figure}
\begin{center}
\mbox{\psfig{file=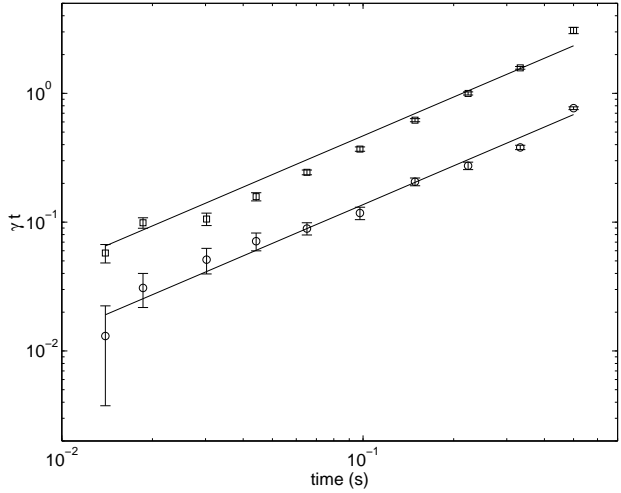,width=3.25in}}
\end{center}
\caption{Determination of the phase damping rates $\gamma_j$.  The circles
represent $j=1$ (proton), and the squares represent $j=2$ (carbon).
$\gamma_3$ is fitted to zero to within experimental uncertainty.}
\label{fig:phasedamp}
\end{figure}

\begin{figure}
\begin{center}
\mbox{\psfig{file=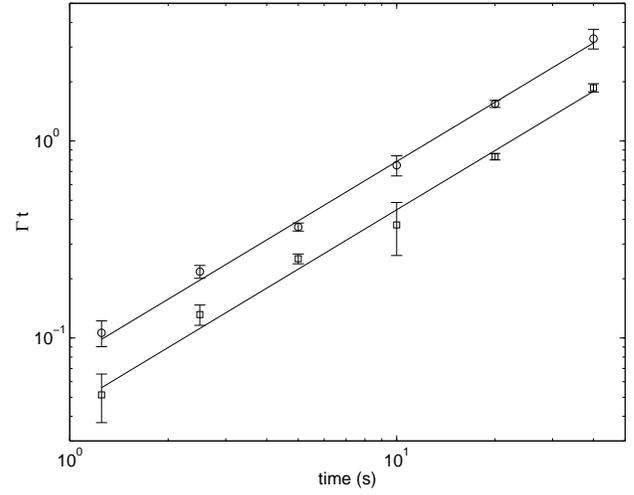,width=3.25in}}
\end{center}
\caption{Determination of the amplitude damping rates $\Gamma_j$.  The
circles represent $j=1$ (proton), and the squares represent $j=2$ (carbon).}
\label{fig:ampdamp}
\end{figure}

Fitting the various rate parameters of $\chi_m$ as a function of time allows
us to characterize the process.  These fits are shown in
Figs.~\ref{fig:jcoupling}--\ref{fig:ampdamp}.  The error bars in these
plots are computed numerically, and are derived solely from the statistics
of the measurements and the fitting procedure.

\begin{table}
\begin{center}
\begin{tabular}{c@{\hspace{16pt}}r@{$\thinspace\pm\thinspace$}l}
$J$                  & $220$  & $3~{\rm Hz}$   \\
$\gamma_1^{-1}$      & $0.73$ & $0.03~{\rm s}$ \\
$\gamma_2^{-1}$      & $0.21$ & $0.01~{\rm s}$ \\
$\Gamma_1^{-1}$      & $12.7$ & $0.3~{\rm s}$  \\
$\Gamma_2^{-1}$      & $22.4$ & $0.7~{\rm s}$
\end{tabular}
\end{center}
\caption{\narrowtext Dynamical time scales for $^{13}{\rm CHCl}_3$, as
measured by QPT.  Recall that the proton is spin 1 and the $^{13}{\rm C}$
nucleus is spin 2.  $\gamma_3$ is fitted to zero to within experimental
uncertainty.}
\label{tab:results}
\end{table}

Our results are summarized in Table~\ref{tab:results}.  In addition to these
data, we found the correlated phase damping rate $\gamma_3$ to be zero to
within statistical precision.  Unfortunately, insufficient precision was
available to determine the amplitude damping temperature parameters $\bar
n_j$.  This is not a significant difficulty, as they are certain to be near
$1 \over 2$ for high temperature systems.  The data are consistent with this
value.

For comparison, the relevant decoherence parameters were measured by
standard techniques.  By the inversion-recovery technique, we found $T_1 =
18.5~{\rm s}$ for the proton and $T_1 = 21.1~{\rm s}$ for carbon.  These
values agree roughly with the time scales $\Gamma_j^{-1}$ for amplitude
damping.  Using the Carr-Purcell-Meiboom-Gill sequence, we found $T_2 =
4.7~{\rm s}$ for the proton and $T_2 = 0.26~{\rm s}$ for carbon.  However,
note that the relevant phase damping time scale is actually $T_2^*$, the
time scale for relaxation due to magnetic field inhomogenity, which can be
measured by fitting the free induction decay.  By this method, we find
$T_2^* = 0.86~{\rm s}$ for the proton and $T_2^* = 0.20~{\rm s}$ for carbon,
values which agree more closely with the time scales $\gamma_j^{-1}$
measured by QPT.

As previously mentioned, the QPT preparation step is nonnegligible for the
controlled-{\sc not} gate fidelity experiment.  However, this is not the
case for the decoherence measurements: the time scales are long compared to
the time to perform the QPT preparation, so small errors in the preparation
are unimportant.  For the experiments at short times to measure the
parameter describing unitary evolution ($J$), the preparation is
nonnegligible, but the operation is simple enough that we are still able to
determine $J$ to within two standard deviations.

\section{Conclusions}

We have demonstrated the implementation of quantum process tomography to
characterize the dynamics of a two-qubit NMR quantum computer.  Such
techniques should prove useful in the future for diagnosing quantum
information processing devices.  However, we should stress that due to the
exponential size of $\chi$, QPT can only be used to characterize the
dynamics of sufficiently small systems.  For large systems, one may only be
able to perform QPT on a small part of the total system, assuming
independence from the rest of the system --- an assumption which can, in
turn, be checked using QPT.

Our measurements of gate fidelities underscore the difficulty of
implementing highly accurate logic gates in NMR.  These fidelity
measurements suggest a per-gate error rate of the order of $10^{-1}$, in
agreement with previous implementations of quantum
algorithms~\cite{Vandersypen}.  It has been suggested that error rates of
$10^{-2}$ can be achieved in solution NMR~\cite{Cory}.  Indeed, if the
previously dicussed cancellation of errors in a long sequence of gates leads
to a $T_2$-limited computation (as observed in~\cite{Vandersypen}), then an
optimistic estimate of $T_2 = 1~{\rm s}$ and $J = 100~{\rm Hz}$ leads to an
approximate error rate of $10^{-2}$.  Nevertheless, even this falls far
short of the requirements for fault-tolerant quantum
computing~\cite{Shor96}: the least stringent estimates indicate a threshold
for fault-tolerance of about $10^{-4}$~\cite{FaultTolThresh}.  Clearly, the
development of fault-tolerant NMR quantum computers will require
substantially modified techniques.  One might construct an alternative model
for fault tolerance, perhaps based on topological properties~\cite{Kitaev}.
However, with a conventional approach to fault tolerance, we will either
need significant improvement of gate fidelity over what is currently
achievable, specialization of fault tolerant protocols to the specific
errors which occur in NMR, or both.

In turn, analyses of the threshold for fault-tolerant quantum computation
typically involve an assumption of independent errors.  Thus, for the
experimental implementation of fault-tolerant computers, it is important to
understand the error model in detail, and in particular, the extent to which
the independence assumption holds.  The results presented in
Section~\ref{sec:decoherence} show that this model {\em is} well-understood
for the $^{13}{\rm CHCl}_3$ NMR system, and furthermore, that the errors are
uncorrelated.  The data are fit well by a combination of phase damping and
independent generalized amplitude damping, with the phase damping
essentially uncorrelated.  Thus there exist simple computational systems in
which an uncorrelated error model is reasonable.

However, note that there are systems which do exhibit correlated
decoherence, namely those that exhibit ``cross-relaxation'' (in the
terminology of NMR).  For example, the simple two-spin proton-carbon system
provided by sodium formate is known to possess
cross-relaxation~\cite{Mayne}.  Investigation of such systems by QPT might
be interesting not only for the purpose of quantum computing, but also for
studying cross-relaxation in the context of conventional NMR.

\section{Acknowledgements}

We thank Constantino Yannoni for sample preparation and Lieven Vandersypen
and Matthias Steffen for experimental assistance.  We also thank all of the
above as well as Mark Sherwood and Xinlan Zhou for numerous enlightening
discussions.  During final preparation of the manuscript, AMC was supported
by the Fannie and John Hertz Foundation.  DWL was partially supported by the
DARPA Ultrascale Program under contract DAAG55-97-1-0341, the IBM Graduate
Fellowship program, and the Nippon Telegraph and Telephone Corporation
(NTT).



\begin{thebibliography}{20}
\bibitem{MultNotation}
  We use multiplicative notation to stand for composition of superoperators.
  In other words, ${\cal E}^{(1)} {\cal E}^{(2)} = {\cal E}^{(1)} \circ
  {\cal E}^{(2)}$.
\bibitem{BetterResultsNote}
  In an earlier experiment, we achieved the fidelities $F_{\sc cnot} = 0.91
  \pm 0.04$, $F_I = 0.93 \pm 0.04$.  Unfortunately, we were unable to
  reproduce these results.
\bibitem{RotFrameNote}
  The time scales $\omega_j^{-1}$ are not relevant because we work in a
  rotating frame, using a pair of local oscilators tuned to match the
  well-known Larmor frequencies.
\end{thebibliography}
\end{document}